\documentclass[aps,prl,twocolumn,groupedaddress,amsmath,amssymb,amsfonts]{revtex4}
\usepackage[dvips]{graphicx}
 \graphicspath{{./pdf_box/}}
\usepackage{subfigure}
\usepackage{latexsym}
\usepackage[english]{babel}
\usepackage{xcolor}
\begin{document}                
\preprint{Submitted to Phys. Rev. Letters}         
\title{Drift-Alfv\'en vortex structures in the edge region of 
a fusion relevant plasma}
\author{N. Vianello, M. Spolaore, E. Martines, R. Cavazzana, G. Serianni, M. Zuin, E. Spada and V. Antoni}
\affiliation{Consorzio RFX, Associazione Euratom-ENEA sulla Fusione, C.so Stati Uniti 4, 35127, Padova, Italy}
\email{nicola.vianello@igi.cnr.it}
\date{\today}

\begin{abstract}
Edge turbulent structures are commonly observed in fusion devices and are generally 
believed to
be responsible for confinement degradation. Among their origin 
Drift-Alfv\'en turbulence is one of the most commonly suggested. Drift-Alfv\'en 
paradigm allows the existence
of localized vortex-like structures observed also in various systems. 
Here we
present the evidence of the presence of drift-Alfv\'en vortices in the edge 
region of RFX-Mod RFP
device,
showing how these structures are responsible for electromagnetic turbulence 
at the edge and
its intermittent
nature.
\end{abstract}
\maketitle
Turbulence represents an outstanding critical issue 
in the physics of magnetically confined plasmas
for thermonuclear fusion research.
Indeed plasma turbulence has been recognized since 
the beginning as the cause of the so-called
\emph{anomalous} particle and energy
transport \cite{Carreras:1997p4071}. In recent years it has been observed that, within incoherent fluctuations,  
coherent structures emerge similar to vortices observed in fluid turbulence \cite{Frisch:1995p3940}. 
These structures have been detected in a variety of devices, ranging from tokamaks  
\cite{Boedo:2001p3938,Fujisawa:2009p3752,Zweben:2007p3942,BenAyed:2009p3835}, 
through stellarators \cite{Grulke:2001p2650}, up to 
reversed field pinches  \cite{Spolaore:2004p245} and linear devices \cite{Grulke:2007p1497}, and 
represent a features shared with astrophysical plasmas \cite{Sundkvist:2005p2076}. Turbulent structures, 
often referred to as \emph{blobs}, are responsible for the high degree of intermittency generally observed. Indeed the 
generation of these structures, arising because of the presence of various instabilities in non-linear regime, is responsible for the 
breaking of self-similarity in the energy cascade process \cite{Carbone:2000p4072}

\emph{Blobs} arising in fusion relevant plasmas, have been extensively studied in the plane perpendicular to the 
main magnetic field \cite{Zweben:2007p3942}, and only recently their 3D features have been experimentally 
addressed \cite{Spolaore:2009p4115}. This interest is enhanced by some analogies with Edge Localized Modes (ELMs), 
which are indeed thought to be associated with parallel current filaments \cite{Kirk:2006p125} . Present theories about blob formation 
and dynamics suggest 
an interchange-like origin, with effects induced by sheath boundary conditions of the material objects intersecting the magnetic 
flux surface \cite{Fundamenski:2007p711}. Plasma quasi-neutrality 
implies the condition $\nabla\cdot\mathbf{J} =0$ on the total current $\mathbf{J}$: considering the  
non vanishing $\nabla_{\perp}$ components of the diamagnetic and polarization current, a parallel current density perturbation 
$\tilde{j}_{\parallel}$ must arise \cite{Krasheninnikov:2008p1282}. Experimental evidence on the existence of filaments associated 
with blobs have been found \cite{BenAyed:2009p3835,Spolaore:2009p4115}. 
Nevertheless, although interchange is believed 
to be responsible for blobs in the Scrape Off Layer plasmas,
they do not represent the only possible mechanism for the generation 
of electromagnetic coherent structures. Drift wave instability, which is thought to dominate plasma 
turbulence in the edge region \cite{Scott:1997p1604}, 
is a nonlinear, non periodic motion
involving disturbances on a background pressure gradient of a magnetized plasma and
eddies of fluid-like motion in which the advecting velocity of all charged species is the
$\mathbf{E}\times\mathbf{B}$ velocity \cite{Scott:2003p1126}. On the theoretical level the electron dynamics is
purely electrostatic if the parallel Alfv\'en transit frequency
is much faster than the thermal electron transit frequency (or equivalently if $\beta \ll m_{e}/M_{i}$) 
and large than any drift-wave frequency. Actually these conditions are 
not satisfied in the edge region of fusion devices\cite{Scott:1997p1604} and the resulting turbulence
and transport level will be determined by
electromagnetic effects in the framework of drift-Alfv\'en dynamics, which represents 
the paradigm for the description of the coupling 
of drift-waves with Kinetic Alfv\'en waves (KAW) \cite{Morales:1997p3845}. 
The key distinction between drift-Alfv\'en dynamics and
magnetohydrodynamics (MHD) is the inclusion of parallel
electron motion
and electron pressure effects. The disturbances in the electric field
arising from the presence of fluid eddies are caused by the tendency of the
electrons to establish a force balance along the magnetic field lines. Pressure disturbances
have their parallel gradients balanced by a parallel electric field,
whose static part is given by a parallel gradient of the electrostatic
potential. Turbulence itself is driven by the background gradient and the electron pressure
and electrostatic potential are coupled together through parallel currents. 
The corresponding magnetic fluctuations could not contribute to a direct enhancing of 
cross-field transport through the so-called magnetic flutter transport \cite{Scott:1997p1604}, but provide an additional coupling 
between parallel drift current and electrostatic drift wave potential \cite{Grulke:2007p1497}. 
As shown in \cite{Scott:1997p1604}
electromagnetic effects are important for drift-wave dynamics
at much smaller values of plasma $\beta$ than expected by pure
MHD considerations. This causes drift-wave dynamics to compete with the
effects of the rather small parallel electron resistivity,
the latter responsible (in the drift-wave framework) for 
the phase relationship between density and potential
fluctuations \cite{Naulin:2003p675}. 

\begin{figure}
%  \centering
  \includegraphics[width=.54\columnwidth]{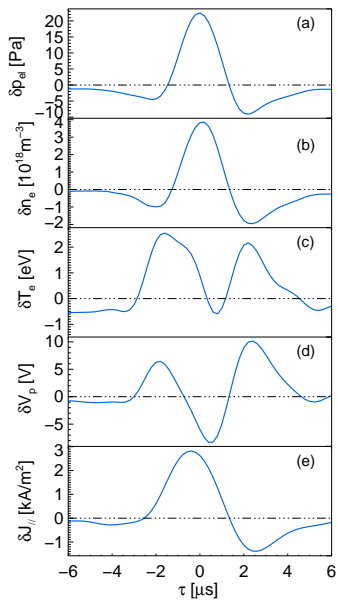}
  \caption{Average coherent structure detected at scale $\tau = 4\mu $s using the electron pressure as reference signal. All 
  waveforms represent variations with respect to average values: (a) electron pressure, (b) electron density, (c) electron temperature, 
  (d) plasma potential, (e) parallel current density}
  \label{fig:fig1}
\end{figure}

In the non-linear regimes, drift-Alfv\'en turbulence may generate non linear structures in the form of electromagnetic 
vortices \cite{Shukla:1986p1196}. 
As aforementioned these structures are generated by the non linear coupling of drift waves and Kinetic Alfv\'en waves. The 
latter exhibit a dispersion relation modified with respect to the shear Alfv\'en waves, with a term including the perpendicular 
wave vector $k_{\perp}$, 
$k_{\parallel}=\frac{\omega/v_{A}}{[1+(k_{\perp}\rho_{s})^{2}]^{1/2}}$ where $v_{A}$ is the Alfv\'en velocity and $\rho_{s}$ 
the ion sound gyroradius $\rho_{s}=c_{s}/\Omega_{i}$. 
These structures have been observed both in astrophysical 
plasmas \cite{Sundkvist:2008p3914,Sundkvist:2005p2076} and in linear devices \cite{Grulke:2007p1497}, but up to now 
they have not been experimentally observed in fusion relevant devices. In this Letter we present a clear experimental 
evidence of the existence of Drift-Alfv\'en-Kinetic (DKA) vortices 
in the edge region of the RFX-mod reversed field pinch experiment \cite{Sonato:2003p2064}. 

\begin{figure}
%  \centering
  \includegraphics[width=.73\columnwidth]{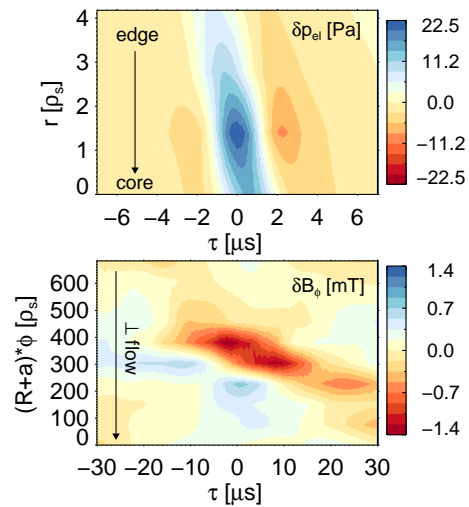}
  \caption{Top: electron pressure as a function of time and radial position normalized 
  to ion sound gyroradius $\rho_{s}$. Bottom: toroidal magnetic field as a function of time and toroidal coordinate normalized 
  to $\rho_{s}$.}
  \label{fig:fig2}
\end{figure}
Despite the peculiar magnetic topology, the edge region of RFP plasmas shares many features with other magnetic 
devices, among which the strong intermittent character of electrostatic fluctuations \cite{Spolaore:2004p245}, 
and the highly sheared $\mathbf{E}\times\mathbf{B}$ flow detected at the edge \cite{Vianello:2005p1976}.
Intermittency manifests itself as a clear departure from self-similarity, and can be imputed to the presence of organized 
structures, \emph{intermittent structures}, which make the process of energy cascade inhomogeneous. They have been extensively 
studied from the experimental point of view, observing their vortex-like shape on the 
perpendicular plane with an associated pressure perturbation \cite{Spolaore:2004p245,Spolaore:2009p4115}. 
They contribute to the cross-field transport for up to 50 \% of the particle losses \cite{Spolaore:2004p245}. Recently 
their electromagnetic features have been experimentally described, revealing the existence of a parallel current 
density fluctuation $\tilde{j}_{\parallel}$ associated to the pressure perturbation, which can represent up to few \% of the 
total parallel current \cite{Spolaore:2009p4115}, but no theoretical explanation has been proposed to determine 
the underlying instability mechanism responsible for the formation of these structures. Although resistive 
interchange are expected to be unstable in RFPs \cite{Merlin:1989p4204},  Drift-Alfv\'en 
turbulence could also play a role, considering the pressure gradient and the $\beta$ condition encountered.

The data presented hereafter have been obtained in the RFX-mod reversed field pinch device ($R/a = 2m/0.459m$ ) 
\cite{Sonato:2003p2064}, operating at relatively low plasma current ($I_{p}\leq 400$kA) and with average density normalized to the 
Greenwald density $n/n_{g}\approx 0.4-0.5$. The typical plasma parameters observed in the edge region for this type 
of discharge are density $n_{e}$ of the order of $1-2\times 10^{19}$ m$^{-3}$, temperature in the range 20-40 eV and magnetic 
field $B_{0}$ around 0.15 T. The corresponding $\beta$ are in the range of 1-2 \% thus ensuring the condition 
$\beta \gg m_{e}/M_{i}$ whereas the typical scale length $\rho_{s}=c_{s}/\Omega_{i}$ is equal to 3-4 mm.
It is worth to remember that in the edge region of RFP plasmas 
the magnetic field is essentially poloidal, so that the perpendicular plane corresponds to the radial-toroidal plane. 

\begin{figure}
%  \centering
  \includegraphics[width=.63\columnwidth]{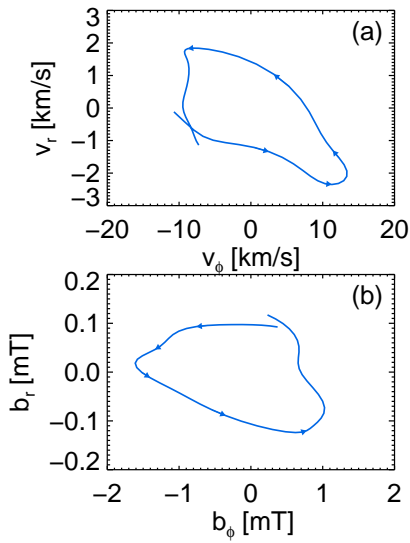}
  \caption{(a) Hodogram of the 
  $\mathbf{E}\times\mathbf{B}$ plasma velocity in the perpendicular plane, showing the closed path corresponding to a 
  parallel vorticity perturbation. (b) Hodogram of the magnetic field fluctuations in the perpendicular plane, showing a closed path 
  corresponding to the transit of a current density filament.}
  \label{fig:fig3}
\end{figure}
A new insertable probe, 
developed in order to study electromagnetic turbulence and described elsewhere \cite{Spolaore:2009p4175}, 
has been used to explore the last 5 cm of the plasma column. 
The system consists of two boron nitride cases, each of them 
housing 5$\times$8 electrostatic pins radially spaced by 6 mm. The pins are used as 5 pins balanced triple probe, 
allowing the simultaneous measurement of plasma density, electron temperature, electron pressure, plasma potential and their 
radial profiles at the same toroidal location, as well as the radial and toroidal components of the $\mathbf{E}\times\mathbf{B}$ 
plasma velocity. The particular arrangement of electrostatic measurements allows a direct estimate of the local fluctuation 
of vorticity $\omega=\nabla\times\mathbf{v}$, where $\mathbf{v}$ is the electric drift 
velocity,  from the floating potential ones $V_{f}$, 
as $\omega_{\parallel}=\frac{1}{B}\nabla_{\perp}^{2}V_{f}$, where plasma potential has been approximated 
by floating potential as usually done 
\cite{Hidalgo:1999p2406}. A radial array of 7 three-axial magnetic coils is located in each case, in 
order to measure the fluctuations of the three components of the magnetic field. Thus a direct estimate of the parallel current
 density can be done from Ampere's law $j_{\parallel}\simeq j_{\theta}=\frac{1}{\mu_{0}}(\partial_{\phi}b_{r}-\partial_{r}b_{\phi})$, 
 virtually in the same toroidal position of the vorticity measurements. 
 \begin{figure}
%  \centering
  \includegraphics[width=.6\columnwidth]{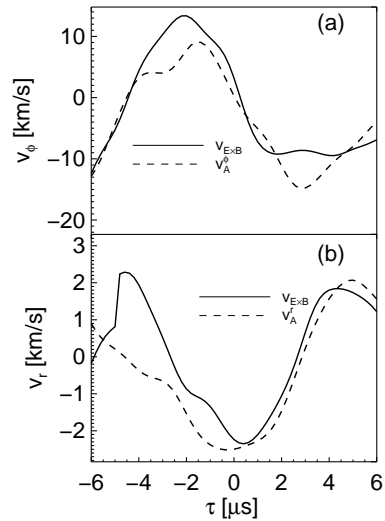}
\caption{Velocity and magnetic field profiles for the average coherent structure detected at $\tau=4\mu$s. (a) 
  Toroidal component of the $\mathbf{E}\times\mathbf{B}$ plasma velocity, and of the corresponding component of 
   Alfv\'en velocity as computed from the 
  magnetic fluctuation measurements. (b) The same of panel (a) but for the radial component. }
 \label{fig:fig4}
\end{figure}
Data were digitally sampled at 5 MHz  with a minimum bandwidth of 700 kHz. The data collected with the insertable probe have 
 been completed with measurements obtained from a toroidal distributed array of magnetic pick-up coils located 
 inside the vacuum vessel, 
 pertaining to the ISIS system \cite{Bolzonella:2003p3766}. The data analysis technique used to disentangle coherent structures 
 from the turbulent background is based on wavelet analysis and has been extensively described elsewhere 
 \cite{Farge:1992p3948}. It allows to locate within the signal the presence of structures at a given temporal scale. 
 This method has been 
 used together with the traditional conditional averaging technique to better extract the common features of the observed structures.

 In figure \ref{fig:fig1} the results of a conditional average procedure are shown. All the time windows used for the average have 
 been chosen using the appearance of an intermittent structure at characteristic time scale $\tau=4 \mu$s (well above 
 the $1/\Omega_{i}$ time scale) on the pressure signal. 
 It can be easily recognized in panel (a) the pressure peak, typical of plasma blobs, mainly determined by 
 electron density (b). The resulting temperature structure display a doubly-peaked pattern, whose impact on the plasma 
 pressure is less important, but contributes in determining the plasma potential pattern, shown in panel (d). 
 It can also be easily recognized that electron density and plasma potential are nearly in phase, 
 with a phase difference 
 around 1 $\mu$s, suggesting the drift-origin of the observed non-linear structure. The typical pattern of 
 current density associated to these measurements is shown in panel (e) of the same figure. It can be easily recognized the existence 
 of a current peak associated to the electron pressure blob. The slight phase shift observed may be imputed to the small 
 deviation of the nominal toroidal position of current measurement with respect to the pressure one, and is consistent with 
 the $\mathbf{E}\times\mathbf{B}$ toroidal propagation of the structure \cite{Spolaore:2009p4175}. The radial array of electron pressure measurements allows to investigate the 
 radial extension of this pressure perturbation, as shown in the upper panel of figure \ref{fig:fig2}. The pressure peak exhibits a radial 
 extent of 2-3 $\rho_{s}$, which is indeed the typical extent of the DKA vortex predicted for example in  
 \cite{Shukla:1986p1196} and observed in \cite{Sundkvist:2005p2076}.
 In the other perpendicular direction the dimension 
 may be estimated using the toroidal distributed array of pick-up coils, and 
 examining the magnetic footprint of the structure. The results of a conditional average procedure still with the appearance of an 
 intermittent structure on electron pressures, is shown in the lower panel of figure \ref{fig:fig2}. In the direction of the flow 
 the structure is larger (of the order of 100 $\rho_{s}$ corresponding to approximately 40 cm). This larger dimension may be imputed 
 to the stretching effect induced by the sheared plasma flow, which is well known to modify vortex structures \cite{Marcus:2000p4073}. 
 It has been already reported elsewhere that these structures are convected by the mean $\mathbf{E}\times\mathbf{B}$ flow and 
 travel in the electron diamagnetic direction \cite{Spolaore:2009p4175}. Using a poloidal array of pick-up coils pertaining to the 
 same ISIS system it is possible to infer the poloidal wave vector $k_{\theta}\approx k_{\parallel}$ 
 of these structures which is of the order of 3.3 m$^{-1}$.

\begin{figure}
%  \centering
  \includegraphics[width=.8\columnwidth]{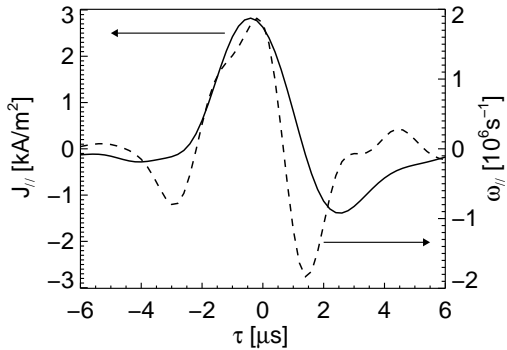}
\caption{Comparison between parallel current density (solid curve) and parallel vorticity (dashed curve) for average coherent structure detected 
  at scale $\tau = 4\mu$s. }
\label{fig:fig5}
\end{figure}
To confirm the 
 vortex nature of the observed structure, the corresponding average $\mathbf{v}_{\mathbf{E}\times\mathbf{B}}$ velocity 
 fluctuations have been calculated. Results are shown in figure \ref{fig:fig3} panel (a) where 
 the hodogram of the drift velocity component on the perpendicular plane is shown. The fluctuating velocities follow a closed 
 trajectory on the perpendicular plane, as a consequence of the vortex-like nature of the observed structure. Similarly the 
 hodogram of the perpendicular components of the magnetic field 
 (figure \ref{fig:fig3} panel (b)) shows a similar closed path, corresponding to 
 the transit of the current density filament observed in figure \ref{fig:fig1} (e). In figure \ref{fig:fig4} panel (a) and (b) 
 the two components of the $\mathbf{E}\times\mathbf{B}$ velocity fluctuations 
 are compared to the corresponding components of Alfv\'en velocity fluctuations as 
 estimated from $\tilde{b}_{\phi}$ and $\tilde{b}_{r}$ and local density measurements, 
 $v_{A}^{r}=\frac{\tilde{b}_{r}}{\sqrt{\rho\mu_{0}}}$ and 
$v_{A}^{\phi}=\frac{\tilde{b}_{\phi}}{\sqrt{\rho\mu_{0}}}$. The good match observed highlights the 
 Alfv\'enic nature of the fluctuating velocities, reinforcing the hypothesis of DKA as underlying physical mechanism.  
 As a final confirmation of the nature of the observed structures we perform 
a direct comparison between vorticity and parallel current density. In the drift-Alfv\'en framework 
electrostatic potential $V_{p}$ and parallel component of the vector potential $A_{\parallel}$ are intrinsically related 
and almost proportional one to the other  \cite{Naulin:2003p675,Liu:1986p3951}. As aforementioned the potential measurement 
arrangement in the U-Probe allows the direct estimate of the vorticity as $\omega_{\parallel}=\frac{1}{B}\nabla_{\perp}^{2} V_{f}$ and 
this can be compared to the parallel component of the current density $\tilde{j}_{\parallel} = \nabla_{\perp}^{2}A_{\parallel}$. This 
comparison is shown in figure \ref{fig:fig5}: the patterns of $\tilde{j_{\parallel}}$ and $\omega$ result from the conditional averaging 
procedure with the same condition used for the previous figure. 
The two quantities are found to be very well correlated one to the other. This last observation together with the drift-type 
phase relation between potential and density, and the alfv\'enicity of the fluctuating velocity,  
establishes without ambiguity the Drift-Kinetic  
nature of the intermittent structures observed at the edge of Reversed Field Pinches.
These measurements suggest the necessity to complete \emph{blob} description with a full electromagnetic characterization, 
and support the theory of Drift-Alfv\'en dynamics as a paradigm for the description of the edge confined region of thermonuclear 
plasmas. \\
%\begin{acknowledgments}
This work was supported by the Euratom Communities under the contract of Association between EURATOM/ENEA. 
The views and opinions expressed herein do not necessarily reflect those of the European Commission.

\end{document}